\documentclass[conference]{IEEEtran}
\IEEEoverridecommandlockouts

\usepackage{cite}
\usepackage{amsmath,amssymb,amsfonts}
\usepackage{algorithmic}
\usepackage{graphicx}
\usepackage{textcomp}
\usepackage{xcolor}

\usepackage{amsmath,amsfonts}
\usepackage{algorithmic}
\usepackage{algorithm}
\usepackage{array}
\usepackage[caption=false,font=normalsize,labelfont=sf,textfont=sf]{subfig}
\usepackage{textcomp}
\usepackage{stfloats}
\usepackage{url}
\usepackage{verbatim}
\usepackage{graphicx}
\usepackage{cite}
\usepackage[all]{nowidow}
\usepackage{xcolor}
\newcommand{\mycomment}[1]{}
\usepackage{pgfplots}
\usetikzlibrary{pgfplots.groupplots}
\usetikzlibrary{shadows,patterns,shapes,arrows,decorations.pathmorphing,backgrounds,positioning,fit,plotmarks,calc,spy}
\def\chartheight{5.5cm}

\def\BibTeX{{\rm B\kern-.05em{\sc i\kern-.025em b}\kern-.08em
    T\kern-.1667em\lower.7ex\hbox{E}\kern-.125emX}}
\begin{document}

\title{Fast Transformer Inference on ARM-Based HMPSoCs\\\thanks{This work was supported by the European Commission, under the CASTOR project; Grant Agreement no. 101167904.}
}

\author{\IEEEauthorblockN{Hang Xu}
\IEEEauthorblockA{\textit{University of Amsterdam} \\
Amsterdam, Netherlands \\
h.xu@uva.nl}
\and
\IEEEauthorblockN{Yixian Shen}
\IEEEauthorblockA{\textit{University of Amsterdam} \\
Amsterdam, Netherlands \\
y.shen@uva.nl}
\and
\IEEEauthorblockN{Thanassis Giannetsos}
\IEEEauthorblockA{\textit{Ubitech} \\
Athens, Greece \\
agiannetsos@ubitech.eu}
\and
\IEEEauthorblockN{Anuj Pathania}
\IEEEauthorblockA{\textit{University of Amsterdam} \\
Amsterdam, Netherlands \\
a.pathania@uva.nl}

}

\maketitle

\begin{abstract}
Transformer models have set new performance standards for Machine Learning (ML) tasks. However, their resource-intensive deployment on resource-constrained edge devices for a cloud-free on-chip transformer inference remains challenging. The {\em ARM Compute Library}~({\em ARM-CL}) framework provides the lowest latency CNN inference on {\em ARM}-based edge devices, but lacks support for transformer inference. In this work, we implement several new transformer kernels in {\em ARM-CL} to  support the transformer natively. Our extended {\em ARM-CL} provides three times faster transformer inference compared to state-of-the-art CPU/GPU implementations on an {\em ARM}-based embedded board.

Furthermore, Heterogeneous Multi-Processor System-on-Chips~(HMPSoCs) powering edge devices imbue them with powerful embedded CPUs and GPUs.  
We introduce cooperative CPU-GPU transformer inference, which executes memory-intensive operations on the CPU while utilizing the GPU for highly parallelizable, compute-intensive operations. This cooperative execution, implemented with minimal overhead, further reduces transformer inference latency by up to 15.72\% compared to the best single-processor inference on {\em ARM-CL}.
\end{abstract}


\section{Introduction}
The transformer architectures initially introduced for Natural Language Processing~(NLP) tasks have rapidly become the foundation for state-of-the-art performance across Machine Learning~(ML) applications, including machine translation~\cite{bahdanau2014neural,shen2025macp,shen2025ssh}, text classification~\cite{garg2020bae}, and recently, vision, audio, and video processing ~\cite{carion2020end,li2026keeping}. 
Performing on-chip transformer inference on edge devices offers improved privacy, responsiveness, reliability, and efficiency. 
However, despite their remarkable functional performance, transformers typically demand billions of floating-point operations and extensive memory bandwidth~\cite{sun2020mobilebert}, which makes them unsuitable for low-latency inference on edge hardware~\cite{shen2022tcps,xiao2022cache,shen2023thermal}. 
Therefore, to integrate transformer-based models seamlessly into daily life, low-latency, on-chip transformer inference is quintessential. 
Consequently, optimizing hardware utilization to achieve efficient transformer inference remains an active research direction.

Heterogeneous Multi-Processor Systems-on-Chip~(HMPSoCs) now underpin most state-of-the-art edge devices~\cite{aghapour2024piqi,shen2026active}, integrating inference-capable embedded CPUs and GPUs. {\em ARM-based} HMPSoCs dominate the mobile platform usage, accounting for the majority of deployments, typically featuring {\em big.Little} asymmetric multi-core CPUs paired with {\em Mali} GPUs. To minimize inference latency, on-chip frameworks must leverage such hardware through custom, hardware-specific kernels. {\em TVM}~\cite{chen2018tvm}, an automated deep learning compiler, facilitates this by providing graph- and operator-level optimizations across diverse hardware backends. However, it underperforms on {\em ARM}-based HMPSoCs. {\em PyTorch}~\cite{imambi2021pytorch}, the popular inference framework for {\em Nvidia}-based HMPSoCs, supports {\em CUDA} but does not work with {\em ARM} Mali GPUs.

In contrast, the developers of the {\em ARM Compute Library~(ARM-CL)}~\cite{sun2017enabling} have tailored it for {\em ARM}-based hardware, providing highly optimized kernels for operations such as Matrix MULtiplications~(MMULs), which are fundamental to transformer models. However, a significant limitation of {\em ARM-CL} is its lack of support for transformers, such as Bidirectional Encoder Representations from Transformers (BERT) and Generative Pre-trained Transformer (GPT), which are more complex than traditional ML architectures like Convolutional Neural Networks~(CNNs).

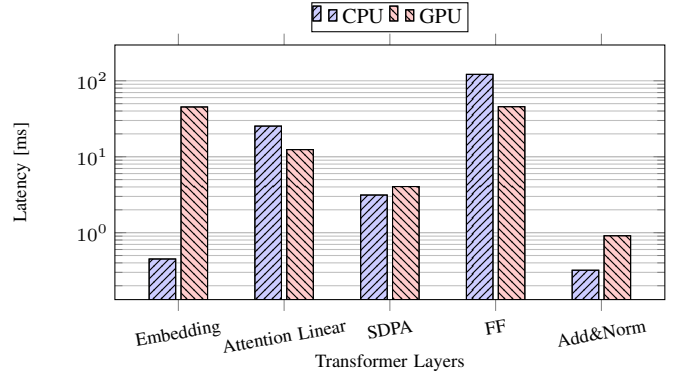
\begin{figure} [t]
\centering
\scriptsize
\begin{tikzpicture}
\begin{axis}[
    width=\columnwidth, height=\chartheight * 0.9, 
    ybar,
    ymode = log,
    log origin=infty,
    ymajorgrids=true, 
    yminorgrids=true,
    enlargelimits=0.15,
    legend style={at={(0.5,-0.15)},
      anchor=north,legend columns=-1},
    ylabel={Latency [ms]},
    xlabel={Transformer Layers},
    xlabel style={yshift=-3pt,},
    symbolic x coords={Embedding, Attention Linear,SDPA, FF, Add\&Norm},
    xtick=data,
    xticklabel style={rotate=10},
    legend style={at={(0.5,1.05)}, anchor=south, cells={anchor=west}, font=\footnotesize, inner sep=0.2pt},
        legend columns=4,
    ]
\addplot [postaction={ pattern=north east lines},fill=blue!20]
	coordinates {
        (Embedding,0.45) 
        (Attention Linear,25.31)
        (SDPA,3.13)
        (FF,121.50)
        (Add\&Norm,0.32)
        };
\addplot [postaction={pattern=north west lines},fill=red!20] 
	coordinates {
        (Embedding,45.23) 
        (Attention Linear,12.41)
        (SDPA,4.05)
        (FF,45.59)
        (Add\&Norm,0.91)
        };
\legend{CPU,GPU}
\end{axis}
\end{tikzpicture}
        \vspace{-0.2cm}
        \caption{Layer-level inference latency comparison between CPU and GPU for {\em BERT-base} transformer with token length $\mathcal{L} = 32 $. Measured on {\em Khadas VIM 3 BASIC} embedded platform with 12\,nm {\em Amlogic A331D} HMPSoC.}
        \label{fig:layer_runtime}
        \vspace{-0.3cm}
\end{figure}

A transformer architecture comprises several different types of layers. The transformer exhibits performance heterogeneity across the HMPSoC processors, with separate layers showing preferences for distinct processors. Therefore, at a higher level, the framework should exploit the performance heterogeneity in HMPSoC processors -- CPU and GPU -- cooperatively for further latency minimization. While there are works that exploit similar performance heterogeneity for CNNs~\cite{aghapour2022cpu}, there is no equivalent work for on-chip transformer inference.

The transformer architecture comprises five primary layers: the Embedding, Attention Linear, Scaled Dot-Product Attention~(SDPA), Feed Forward~(FF), and Add \& Normalization~(Add\&Norm) layers. 
Figure \ref{fig:layer_runtime} shows that the CPU achieves better latency for the Embedding, SDPA, and Add\&Norm layers than the GPU. 
In particular, the SDPA layer demonstrates a significant latency advantage on the CPU compared to the GPU.
Conversely, the GPU outperforms the CPU for the Attention Linear and FF layers.
This result is unique to embedded CPUs and GPUs~\cite{al2020novel}. Figure~\ref{fig:layer_runtime} underscores the potential for a hybrid CPU-GPU execution strategy in embedded systems, where workloads are distributed based on the computational characteristics of each layer. This distribution leads to more efficient transformer inference across HMPSoCs.

\begin{figure*}[ht]
\centering
\includegraphics[width=0.9\linewidth]{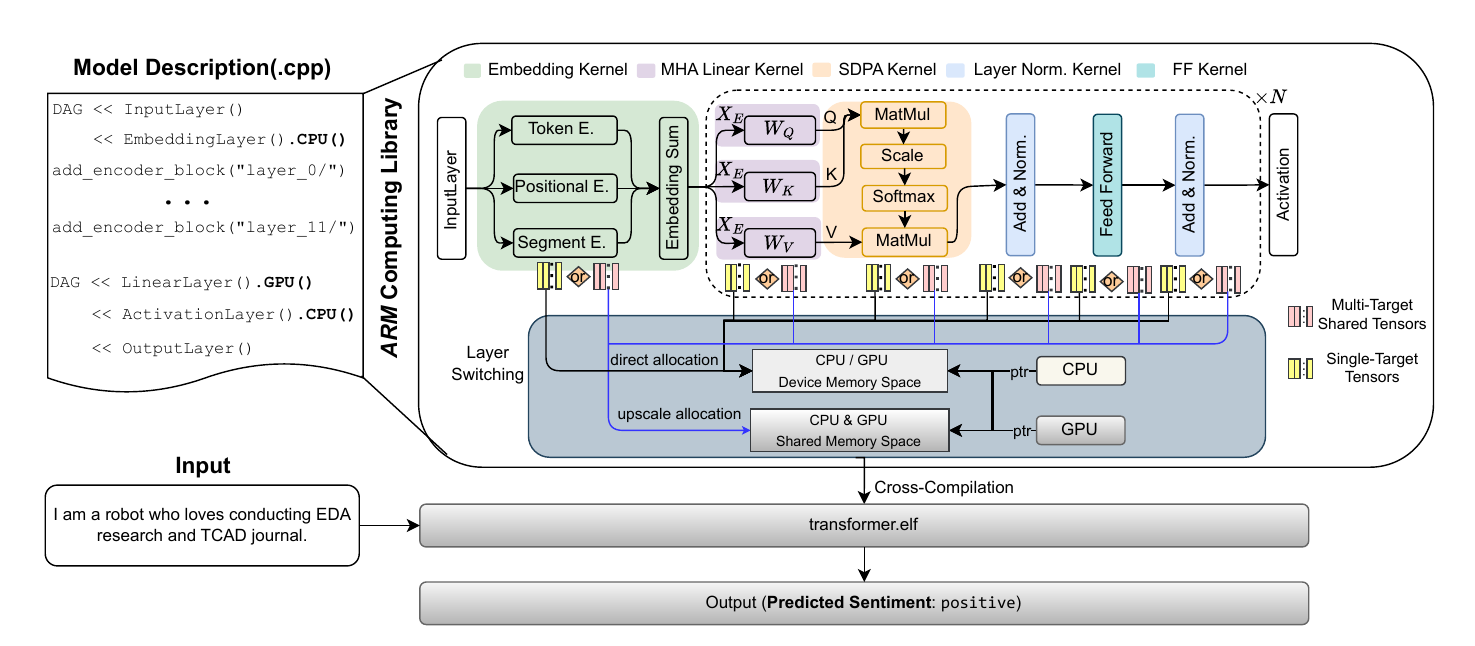}
    \caption{{\em ARM-CL} based implementation for CPU-GPU layer-switched transformer inference using {\em BERT-base}. Model description outlines the model structure, layer connectivity, and specific computing kernels, which are compiled into a model-specific executable. Shared tensors between the CPU and GPU kernels for layer-switching communication are set up during configuration phase.} 
    \label{fig:armcl_extend}    
    \vspace{-0.5cm}
\end{figure*}

\textbf {Our Novel Contributions:} We make the following original contributions within the purview of this work.

\begin{itemize} 
\item We extend the {\em ARM-CL} framework using state-of-the-art ML kernels to enable on-chip transformer inference on embedded CPUs and GPUs in {\em ARM}-based HMPSoCs.

\item We perform a micro-benchmark-based characterization of transformer execution across HMPSoC processors, providing the first quantitative analysis of layer-wise performance heterogeneity.

\item We demonstrate a substantial reduction in inference latency using CPU-GPU layer switching on a {\em Khadas VIM 3} embedded platform with an {\em Amlogic A331D} HMPSoC.

\end{itemize}

\textbf {Open-Source Contributions}: The code for the ARM-CL extensions proposed in this work are publicly available at https://github.com/dondavan/fast-transformer-on-acl under {\em MIT} license.

\section{Related Work}

Deploying transformer models on resource-constrained edge devices has drawn increasing attention due to the growing demand for low-latency, high-accuracy ML applications.
Despite their impressive functional performance, transformers introduce significant computational challenges compared to CNNs, being 1.58x to 41x slower in inference on edge~\cite{wang2022towards}.

\textbf{Optimizing Transformer Architecture and Model Compression.}
Efforts to improve the efficiency of transformer models include architecture-level optimizations. Researchers have developed \textit{MobileBERT}~\cite{sun2020mobilebert} and \textit{SqueezeBERT}~\cite{iandola2020squeezebert} to maintain the accuracy of {\em BERT} while significantly reducing inference latency and model size. 
These approaches focus on creating denser, more efficient architectures by introducing specialized neural micro-architectures or using group convolution layers, achieving speedups of 4.3x to 5.5x on mobile platforms compared to {\em BERT-base}~\cite{sun2020mobilebert, iandola2020squeezebert}. Furthermore, researchers have also explored compression techniques such as network pruning and quantization~\cite{han2015deep} to reduce the size of transformers. 
{\em DistilBERT}~\cite{sanh2019distilbert}  uses knowledge distillation to reduce model size while preserving 97\% of the original performance, resulting in a 60\% faster run-time.
These methods focus on minimizing model complexity but often face trade-offs in terms of accuracy, especially when deployed in resource-constrained environments.

\textbf{Hardware-Specific Optimizations for Inference.}
Optimizing transformer inference for specific hardware platforms, such as {\em ARM}-based HMPSoCs, has garnered significant attention. \textit{TVM}~\cite{chen2018tvm} is a widely used automated deep-learning compiler that supports various hardware backends and applies graph- and operator-level optimizations. However, {\em TVM}’s general-purpose approach limits its ability to fully exploit hardware-specific features, such as those in {\em ARM} architectures. Additionally, the popular {\em PyTorch}~\cite{imambi2021pytorch} framework lacks support for {\em ARM Mali} GPUs. Conversely, hardware-specific libraries like \textit{ARM-CL}~\cite{sun2017enabling}  provide optimized kernels for matrix multiplications and convolutions but lack support for the specialized operations required by transformer models, such as self-attention and layer normalization.

\textbf{Dynamic Layer-Wise Optimization Approaches.} Recent works explore cooperative CPU-GPU CNN inference for {\em ARM}-based~\cite{karatzas2023omniboost,aghapour2022cpu} and {\em Nvidia}-based~\cite{taufique2025hidp, taufique2025twill} embedded platforms. However, the direct applicability of these works to transformer inference in ARM-based platforms remains questionable. Authors of~\cite{chang2023pipebert} explore parallelizing the computation of transformer inference across ARM-based HMPSoC processors, but limit themselves to CPU cores. Therefore, distributing transformer inference computation across ARM-based HMPSoC processors has not been thoroughly explored.

{

\textbf{Deep Learning Accelerators (DLAs).} DLAs are domain-specific hardware platforms such as Neural Processing Units~(NPUs), designed with fixed-function units to support deep learning operations. Several prior works have explored running edge inference on DLAs~\cite{kim2020autoscale, dagli2024shared}. DLAs provide specialized acceleration for machine learning workloads. However, a major limitation for DLAs is its lack of programmability and support for only a limited set of operations required for transformer inference ~\cite{jeong2022band}. Consequently, existing work targeting DLAs has largely focused on CNN workloads~\cite{karatzas2023omniboost}.

\section{Single-Processor Transformer Inference} \label {Sec: Arm_cl_extend}

The current {\em ARM-CL} framework supports only CNN inference. 
Since transformer inference requires additional operations, we have enhanced {\em ARM-CL} to provide native support for transformer-based models on HMPSoCs. Our extensions encompass new custom kernels for embedding, Multi-Head Attention (MHA), SDPA, and layer normalization in {\em ARM-CL} to enable various transformer layers. As such, these kernels are absent from {\em ARM-CL}. Figure~\ref{fig:armcl_extend} uses abstraction to show the on-chip transformer inference using the {\em BERT-base} transformer enabled by these kernels.

The kernel implementation is computationally equivalent to the state-of-the-art approach used in {\em PyTorch}~\cite{imambi2021pytorch} on desktop platforms, with optimizations targeting edge devices. It is intended to report state-of-the-art transformer inference performance on edge devices. Unlike {\em TVM} or {\em ExecuTorch}, which compile the model together with supplementary operations into a single runtime executable, {\em ARM-CL} provides low-level control over runtime execution. This enables closer adherence to the computation model and facilitates runtime performance analysis with minimized overhead.

These kernels leverage {\em NEON} intrinsics and {\em Open Computing Language (OpenCL)} to accelerate execution on the CPU and GPU, respectively. NEON intrinsics enable Single Instruction, Multiple Data~(SIMD) execution by utilizing 128-bit registers on the CPU. The GPU hardware architecture and execution model differ fundamentally from those of the CPU. Consequently, GPU kernels employ computation strategies distinct from CPU kernels, such as tiled MMUL. These parallel computation techniques enhance kernel efficiency and enable more effective utilization of the underlying hardware.

We use these kernels to port five different transformer architectures to {\em ARM-CL}: {\em BERT-base}, {\em DistilBERT}, {\em MobileBERT}, {\em SqueezeBERT}, and {\em GPT-2}.

\section{Transformer Inference Characterization} \label {Sec: Characterization}

\begin{figure*}[ht]
\begin{tikzpicture}
\begin{groupplot}[
    group style={
        group name=my plots,
        group size=3 by 3,
        ylabels at=edge left
    },
    footnotesize,
    width=0.33\textwidth, 
    height=\chartheight *0.8,
    tickpos=left,
    xtick=data,
    ytick align=outside,
    xtick align=outside,
    enlarge x limits=true,
    legend style={at={(-0.7,2.6)},
      anchor=south,legend columns=-1}
]

\nextgroupplot[title={$\mathcal{L}=16$},ylabel=$T_{CPU/GPU}$]
\addplot[solid] coordinates{
         (192,1)
         (384,1)
         (576,1)
         (768,1)
         (960,1)
        };

\addplot[mark=Mercedes star flipped,
         mark options={solid},
         style={dotted},
         color={cyan}] coordinates{
         (192,0.018389146166171103)
         (384,0.012230776216217633)
         (576,0.009168789353163672)
         (768,0.00832281938617308)
         (960,0.007390059919701914)
        };
\addplot[mark=Mercedes star,
         mark options={solid},
         style={densely dotted},
         color={magenta}] coordinates{
         (192,1.6552616048227191)
         (384,2.9505965243427363)
         (576,2.4094076971308183)
         (768,3.1956998128630505)
         (960,3.2005175386947586) 
        };
\addplot[mark=x,
         mark options={solid},
         style={loosely dotted},
         color={purple}] coordinates{
         (192,0.5651088685635685)
         (384,0.5891580311187644)
         (576,0.6838726409611701)
         (768,0.7806028430382653)
         (960,0.825019848203464)
        };
\addplot[mark=+,
         mark options={solid},
         style={densely dashdotted},
         color={teal}] coordinates{
         (192,4.598904120823987)
         (384,4.498927868033486)
         (576,3.629046362128536)
         (768,4.461504824436807)
         (960,4.452859912064388)
        };
\addplot[mark=star,
         mark options={solid},
         style={dashed},
         color={violet}] coordinates{
         (192,0.35313649052161317)
         (384,0.31765437813188285)
         (576,0.30150062833866004)
         (768,0.3083740123782698)
         (960,0.30409180909825057)
        };

\nextgroupplot[title={$\mathcal{L}=64$}]
\addplot[solid] coordinates{
         (192,1)
         (384,1)
         (576,1)
         (768,1)
         (960,1)
        };
\addplot[mark=Mercedes star flipped,
         mark options={solid},
         style={dotted},
         color={cyan}] coordinates{
    (192,0.026050933491659305)
    (384,0.018192683937699027)
    (576,0.014213049000130628)
    (768,0.014377380811003359)
    (960,0.014698843042652733)
};
\addplot[mark=Mercedes star,
         mark options={solid},
         style={densely dotted},
         color={magenta}] coordinates{
    (192,1.0814211476933542)
    (384,1.6240597699105233)
    (576,1.2764406824071357)
    (768,1.2850909378837232)
    (960,1.1118939736035078)
};
\addplot[mark=x,
         mark options={solid},
         style={loosely dotted},
         color={purple}] coordinates{
    (192,0.6560674374117186)
    (384,0.7239897821793376)
    (576,0.7188857739762621)
    (768,0.680868255233455)
    (960,0.679133074531269)
};
\addplot[mark=+,
         mark options={solid},
         style={densely dashdotted},
         color={teal}] coordinates{
    (192,1.9359256554316613)
    (384,1.7773248594150663)
    (576,1.5311494493358713)
    (768,1.5980019010649695)
    (960,1.4225323726623746)
};
\addplot[mark=star,
         mark options={solid},
         style={dashed},
         color={violet}] coordinates{
    (192,0.3414686170218304)
    (384,0.3520296767447331)
    (576,0.36453290527682464)
    (768,0.40338642026597366)
    (960,0.39031761753016175)
};

\nextgroupplot[title={$\mathcal{L}=128$}]
\addplot[solid] coordinates{
         (192,1)
         (384,1)
         (576,1)
         (768,1)
         (960,1)
        };

\addplot[mark=Mercedes star flipped,
         mark options={solid},
         style={dotted},
         color={cyan}] coordinates{
    (192,0.02899933661865962)
    (384,0.02309811958487604)
    (576,0.0235726066771981)
    (768,0.020844121675773965)
    (960,0.02035466574571435)
};
\addplot[mark=Mercedes star,
         mark options={solid},
         style={densely dotted},
         color={magenta}] coordinates{
    (192,0.8230507209585235)
    (384,1.1569305993594452)
    (576,0.8179691964614808)
    (768,0.8066429708967532)
    (960,0.7047930778835494)
};
\addplot[mark=x,
         mark options={solid},
         style={loosely dotted},
         color={purple}] coordinates{
    (192,0.7710738565829585)
    (384,0.6996839649244283)
    (576,0.6948719248850963)
    (768,0.5764221185250095)
    (960,0.5889426830142215)
};
\addplot[mark=+,
         mark options={solid},
         style={densely dashdotted},
         color={teal}] coordinates{
    (192,1.0736667796634825)
    (384,1.0235801947796215)
    (576,0.9083318098390554)
    (768,0.912975248681398)
    (960,0.8129049998975757)
};
\addplot[mark=star,
         mark options={solid},
         style={dashed},
         color={violet}] coordinates{
    (192,0.3537065153869493)
    (384,0.38378603723609955)
    (576,0.4062061767937139)
    (768,0.40858762583992336)
    (960,0.39223864316881984)
};
\nextgroupplot[title={$\mathcal{L}=256$},ylabel=$T_{CPU/GPU}$]

\addplot[solid] coordinates{
         (192,1)
         (384,1)
         (576,1)
         (768,1)
         (960,1)
        };

\addplot[mark=Mercedes star flipped,
         mark options={solid},
         style={dotted},
         color={cyan}] coordinates{
    (192,0.04260762649319085)
    (384,0.037525990470163716)
    (576,0.033650914284014365)
    (768,0.027862764976017855)
    (960,0.030873229381321785)
};
\addplot[mark=Mercedes star,
         mark options={solid},
         style={densely dotted},
         color={magenta}] coordinates{
    (192,0.742490109808399)
    (384,0.8898465859730507)
    (576,0.5895610664839401)
    (768,0.5543645828522654)
    (960,0.500591968894836)
};
\addplot[mark=x,
         mark options={solid},
         style={loosely dotted},
         color={purple}] coordinates{
    (192,0.8801618746243023)
    (384,0.6277349777749357)
    (576,0.5650156188330305)
    (768,0.49259836245358535)
    (960,0.4281664672225567)
};
\addplot[mark=+,
         mark options={solid},
         style={densely dashdotted},
         color={teal}] coordinates{
    (192,0.8912572624158567)
    (384,0.7729430229833023)
    (576,0.6751728677569209)
    (768,0.6067243388099794)
    (960,0.5445306389286232)
};
\addplot[mark=star,
         mark options={solid},
         style={dashed},
         color={violet}] coordinates{
    (192,0.3453286592406958)
    (384,0.3585879613668385)
    (576,0.35109089057182724)
    (768,0.3555671421667102)
    (960,0.33968654411900423)
};

\nextgroupplot[title={$\mathcal{L}=384$}, xlabel={$d_{model}$ [floats]}]

\addplot[solid] coordinates{
         (192,1)
         (384,1)
         (576,1)
         (768,1)
         (960,1)
        };

\addplot[mark=Mercedes star flipped,
         mark options={solid},
         style={dotted},
         color={cyan}] coordinates{
    (192,0.06239273627872793)
    (384,0.050802329213432765)
    (576,0.04413003953657106)
    (768,0.038905152108844855)
    (960,0.04948987981121532)
};
\addplot[mark=Mercedes star,
         mark options={solid},
         style={densely dotted},
         color={magenta}] coordinates{
    (192,0.7275630979113041)
    (384,0.8487038886335844)
    (576,0.5413225391971387)
    (768,0.4818834668327298)
    (960,0.44645724080211285)
};
\addplot[mark=x,
         mark options={solid},
         style={loosely dotted},
         color={purple}] coordinates{
    (192,0.7296485293489192)
    (384,0.6251537613549525)
    (576,0.49925664399220665)
    (768,0.4456312328459274)
    (960,0.4055791226390876)
};
\addplot[mark=+,
         mark options={solid},
         style={densely dashdotted},
         color={teal}] coordinates{
    (192,0.5822649539656756)
    (384,0.6262128409996878)
    (576,0.5776927843945981)
    (768,0.518744577326564)
    (960,0.47940891158514826)
};
\addplot[mark=star,
         mark options={solid},
         style={dashed},
         color={violet}] coordinates{
    (192,0.34040232555807637)
    (384,0.3711162031328234)
    (576,0.3392041767775287)
    (768,0.33948959242770216)
    (960,0.33555491022082967)
};
\nextgroupplot[title={$\mathcal{L}=512$}]

\addplot[solid] coordinates{
         (192,1)
         (384,1)
         (576,1)
         (768,1)
         (960,1)
        };
        
\addplot[mark=Mercedes star flipped,
         mark options={solid},
         style={dotted},
         color={cyan}] coordinates{
    (192,0.07474425780245342)
    (384,0.054279676784114994)
    (576,0.043145880661242135)
    (768,0.05821248272050492)
    (960,0.050033296630270736)
};
\addplot[mark=Mercedes star,
         mark options={solid},
         style={densely dotted},
         color={magenta}] coordinates{
    (192,0.802908089189778)
    (384,0.23151028902363327)
    (576,0.5114245053868243)
    (768,0.49110478263748136)
    (960,0.41472928171573037)
};
\addplot[mark=x,
         mark options={solid},
         style={loosely dotted},
         color={purple}] coordinates{
    (192,0.6874585390144459)
    (384,0.5871774485584721)
    (576,0.5073108519406347)
    (768,0.4193843843823482)
    (960,0.1815663647671595)
};
\addplot[mark=+,
         mark options={solid},
         style={densely dashdotted},
         color={teal}] coordinates{
    (192,0.5319339431552759)
    (384,0.4515010986802552)
    (576,0.40944648446857435)
    (768,0.5016244364812785)
    (960,0.4227382516271749)
};
\addplot[mark=star,
         mark options={solid},
         style={dashed},
         color={violet}] coordinates{
    (192,0.3364102062545882)
    (384,0.34787264697150694)
    (576,0.3192263058300709)
    (768,0.3291724481141878)
    (960,0.2697235089864416)
};

\legend{Reference $T_{CPU} = T_{GPU}$, Embedding, Attention Linear, SDPA, FF, Add\&Norm}
\end{groupplot}
\end{tikzpicture}
    \vspace{-0.3cm}
    \centering
    \caption{Exploring $T_{CPU/GPU}$ of model depth $d_{model}$ with different token length $\mathcal{L}$. Measured using {\em BERT-base} transformer on {\em Khadas VIM 3 BASIC} with 12\,nm {\em Amlogic A331D} HMPSoC. }
    \label{fig:layer_runtime_ratio}        
    \vspace{-0.3cm}
\end{figure*}
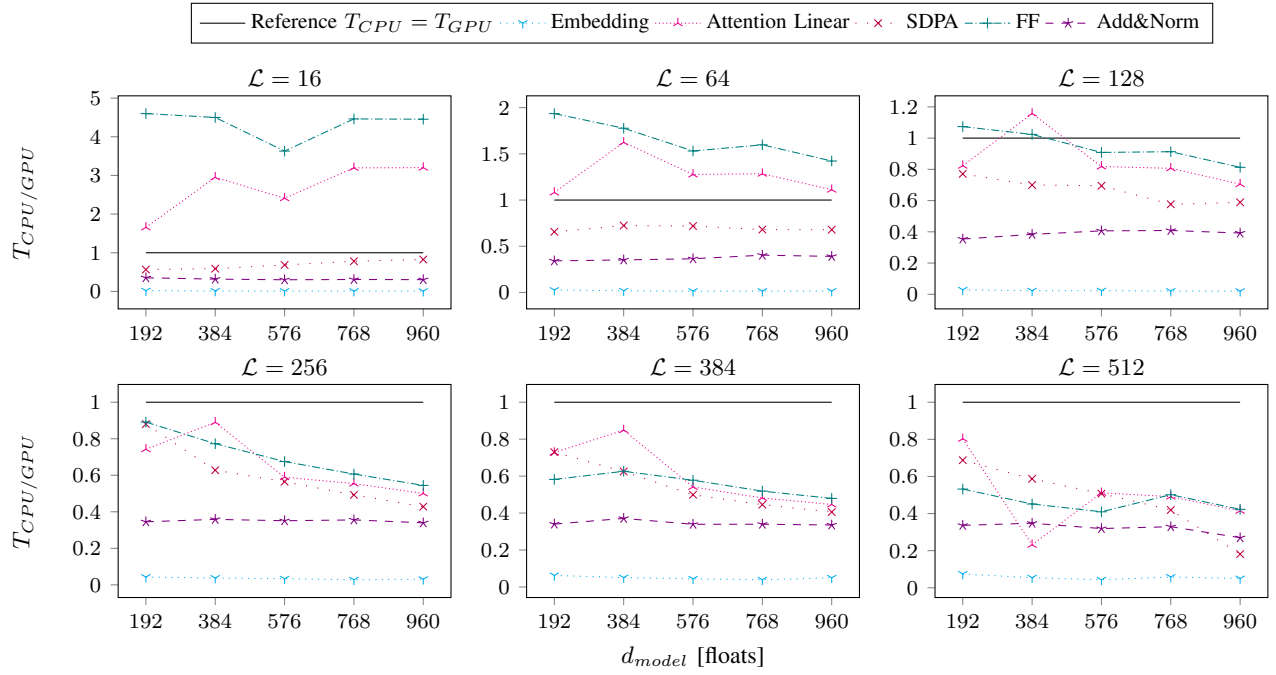

We perform a detailed characterization of on-chip transformer inference using our {\em ARM-CL} extensions on an {\em ARM}-based HMPSoC. 
{\em ARM-CL} offers low-level control over the computing workload while introducing minimal supplementary overhead.
The characterization uses micro-benchmarking with synthetic multi-size transformer layers to explain the performance heterogeneity of different kernels on various HMPSoC processors, as seen in Figure~\ref{fig:layer_runtime}.  We analyze, in isolation, all five fundamental transformer layers: the Embedding, Attention Linear, SDPA, FF, and Add\&Norm layers. The layer runtime characterization is applicable to all transformer-based models, as these layers are common across transformer family. Thus, our layer characterization analysis is model-agnostic.

Let $\mathcal{L}$ denote the input sequence length in tokens, $\mathcal{L}$ varies in value from 1 to an architecture-defined maximum. For example, {\em BERT-base} supports sequences up to 512 tokens. Another key parameter is the hidden dimension size, denoted as $d_{model}$, which defines the tensor width and remains constant across all transformer layers. 
While $d_{model}$ and $\mathcal{L}$ are the two primary parameters that determine the computational workload, the runtime input sequence length $\mathcal{L}$ does not influence model accuracy.
{\em BERT-base} uses $d_{model} = 768$ floats. {\em GPT-2} supports sequences of up to 1024 tokens and uses $d_{model} = 768$. Let $T_{CPU}$ and $T_{GPU}$ represent the execution time of a layer on the CPU and GPU, respectively. The ratio $T_{CPU/GPU}$ indicates their relative performance; a higher ratio suggests the layer executes faster on the GPU.

Figure ~\ref{fig:layer_runtime_ratio} illustrates the variation in $T_{CPU/GPU}$ per different transformer layers as $d_{model}$ increases from 192 to 960 floats and the $\mathcal{L}$ ranges from 16 to 512 tokens. The reference line at $T_{CPU/GPU} = 1$ indicates equal execution time on CPU and GPU.
When $\mathcal{L} \geq 256 $, $T_{CPU/GPU} < 1$ for all layer types, regardless of the value of $d_{model}$. 
We will focus only on computational workload analysis with  $\mathcal{L} \leq 128 $ in this characterization, where CPU and GPU have comparative advantages in execution. In the rest of the section, we present a more detailed analysis of GPU performance bottlenecks arising from hardware limitations when $\mathcal{L} \geq 128 $.

\textbf {Embedding Layers}. 
The Embedding layer reads in a set of pre-trained model weights to map individual input tokens to their corresponding vector representations. This process is predominantly memory-bound, involving extensive memory accesses and minimal computation. 
As CPUs are optimized for efficient serial operations, they demonstrate superior performance on Embedding layers. This observation is consistent across all tested values of $\mathcal{L}$ and $d_{model}$.

\textbf {Attention Linear Layers}. 
The Attention Linear layers consist of three linear transformations applied to Query, Key, and Value tensors using MMULs. 
Each MMUL operates between an input tensor of shape $(d_{model},d_{model})$ and a weight tensor of shape $(d_{model},d_{model})$, resulting in a total of $3 \cdot 2 \cdot \mathcal{L} \cdot d_{model}^2$ FLoating-point Operations (FLOPs) across the three MMULs. Consequently, the computational cost of these MMUL operations scales with both $\mathcal{L}$ and $d_{model}$.
While GPUs initially outperform CPUs due to their highly parallel execution model and the use of tiled MMULs in GPU kernels, GPU performance degrades significantly as the MMUL size increases. We attribute this degradation primarily to memory access overheads.

\textbf {SDPA Layers}.  
The SDPA layer processes Query, Key, and Value tensors through a memory-intensive reshape and permutation. Input tensors with shape $(\mathcal{L},d_{model})$ are transformed into $(\frac{d_{model}}{h},h,\mathcal{L})$,
followed by compute-intensive MMUL operations. Specifically, two MMULs are performed between the Query and Key heads and the resulting attention scores and Value heads, which have $4 \cdot \mathcal{L}^2 \cdot d_{model}$ FLOPs. 
The MMUL complexity of the SDPA layer is lower than that of the attention linear and FF layers, as its computational complexity scales quadratically with $\mathcal{L}$. In addition, tensor initialization and element-wise addition operations occur in GPT-2. The memory-bound phases favor CPU execution, whereas the MMUL phases benefit from GPU parallelism. Overall, SDPA performance is comparable between the CPU and GPU.

\textbf {FF Layers.} 
The FF layer consists of two MMUL operations separated by a memory-intensive activation operation with a relatively small workload. The first MMUL transforms the input tensor of shape $(\mathcal{L},d_{model})$ into a higher-dimensional space by multiplying it with a weight tensor of shape $(d_{model},d_{ff})$, where $d_{ff}$ is typically larger than $d_{model}$ to project learnable space (e.g., $d_{ff} = 3072$ for {\em BERT-base}). After the activation, projected tensor is transform back into $d_{model}$ dimension by multiplying with a weight tensor of shape $(d_{ff},d_{model})$. In total, the two MMULs require $4 \cdot \mathcal{L} \cdot d_{model} \cdot d_{ff}$ FLOPs. Consequently, the FF layer benefits from GPU parallelization. However, similar to Attention Linear layers, these gains diminish as MMUL sizes grow and exceed the GPU’s optimal memory and compute balance. The FF layer has higher MMUL complexity compared to SDPA, but it benefits from fewer memory operations, resulting in greater GPU speedup, as reflected by a higher $T_{CPU/GPU}$.

\textbf {Add\&Norm Layer.} 
The Add\&Norm layer performs element-wise addition followed by layer normalization, both of which involve frequent memory accesses with minimal computation. These accesses make the layer highly memory-bound.  Similar to Embedding layers, the CPU consistently outperforms the GPU due to its superior efficiency in handling memory-intensive, low-compute tasks.

\section{Multi-Processor Transformer Inference}

\begin{figure}[t]
\centering
 \includegraphics[width=0.9\linewidth]{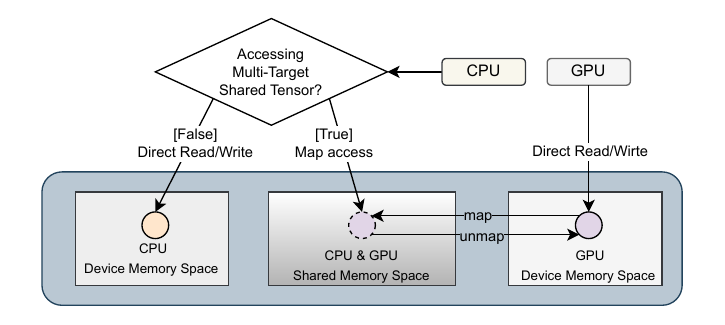}
    \vspace{-0.4cm}
    \caption{CPU-GPU interactions during multi-processor transformer inference.}    
    \label{fig:CPU-GPU_interact}
    \vspace{-0.3cm}
\end{figure}

Our characterization in Section~\ref{Sec: Characterization}   demonstrates the performance heterogeneity of transformer layers across different HMPSoC processors. This observation suggests that we can reduce the transformer inference latency by adopting a CPU-GPU layer-switched execution strategy, assigning each layer to the processor where it runs most efficiently.

{\em ARM-CL} translates neural architectures into model graphs before scheduling their execution. The authors of~\cite{aghapour2022cpu} applied graph partitioning to divide CNN model graphs into subgraphs, which they assign to different HMPSoC processors. Inter-subgraph communication was handled by inserting \texttt{Sender} and \texttt{Receiver} nodes that transferred data via \texttt{memcpy} operations. However, transformer architectures introduce additional complexity due to recurrence and residual connections, resulting in a more intricate dataflow graph. We observed excessive recursive \texttt{memcpy} traffic between the \texttt{Sender} and \texttt{Receiver} nodes, when applying the method from~\cite{aghapour2022cpu} to layer-switched transformer inference.

Mobile HMPSoC architectures, such as {\em Amlogic A331D} support a shared memory space with the host, enabling the CPU and GPU to exchange data without explicit \texttt{memcpy} operations. This work leverages this capability to reduce the communication overhead associated with a cooperative CPU-GPU transformer inference approach. 

We propose a novel CPU-GPU layer-switched transformer inference, illustrated in Figure~\ref{fig:armcl_extend}. Our approach distributes layer computation to the processor that delivers the best performance. We use single-target tensors for consecutive layers on the same processor, while we employ multi-target shared tensors at CPU-GPU transitions, as depicted in Figure~4.
To address the heterogeneous memory architectures of the CPU and GPU, we adopt an ``upscale'' approach by promoting CPU tensors to GPU tensors. We allocate shared tensors in a GPU memory region that can be directly accessed by the CPU via {\em OpenCL}, without explicit \texttt {memcpy} operations. 

\section{Experimental Evaluation}

We use a {\em Khadas VIM 3 BASIC} board for our evaluations. The board features a 12\,nm {\em Amlogic A331D} HMPSoC. It contains an {\em ARM big.Little} asymmetric multi-core CPU with a quad-core high-performance, high-power {\em ARM Cortex-A73} {\em big} CPU cluster running at 2.2\,GHz and a dual-core low-performance, low power {\em ARM Cortex-A53} {\em Little} CPU cluster running at 1.8\, GHz. Given the high-performance emphasis of this work, we only employ the high-performance {\em big} CPU. The HMPSoC also contains a quad-core {\em ARM Mali G52} running at 0.8\,GHz. 
The operating system is {Ubuntu 22.04.4} with {\em Linux} kernel v4.9. We use {\em ARM-CL} v21.02 with {\em OpenCL 3.0} as the baseline version to perform the extensions.

On resource-constrained embedded platforms, available memory is typically limited, with DRAM capacity often restricted to 2–4\,GB. We evaluated models with varying parameter sizes and found that {\em BERT-base} (approximately 300\,MB) runs without issues, whereas deploying {\em BERT-large} (approximately 1\,GB) resulted in runtime crashes. Furthermore, cache size has a significant impact on runtime performance, and HMPSoC components exhibit distinct architectural characteristics. On the evaluated platform, the CPU is equipped with a 2\,MB L2 cache, while the GPU has a 128\,KB L2 cache. Similar trends are observed across other embedded hardware platforms, where GPUs typically have much smaller L2 caches—often only a few kilobytes—compared to CPUs, which usually provide several megabytes of L2 cache.

Figure~\ref{fig:layer_runtime_ratio} illustrates the impact of memory characteristics on inference latency. We observe that the GPU performance advantage for executing the Attention Linear and FF layers diminishes for sequence lengths $\mathcal{L} \geq 128 $. These layers perform matrix MMULs between the input matrix from the previous layer (cached) and the weight matrix stored in DRAM. As observed from the graph, in the case of $\mathcal{L} = 128, d_{model} = 384$, the input matrix size is $4 * 128 * 384 = 192$\,KB, which marks a threshold where larger MMUL workloads reduce cache efficiency and, consequently, increase latency.

\textbf{Benchmarks.} We use five different transformer architectures -- {\em BERT-base}, {\em DistilBERT}, {\em MobileBERT}, {\em SqueezeBERT}, and {\em GPT-2}  -- in this work.
Unless stated otherwise, we conduct all experiments using each model's default hidden dimension ($d_{model}$) and a fixed input token length of 32. 
We implement the BERT-base~\cite{devlin2018bert} model from scratch in {\em ARM-CL}, which we then use to derive other transformer models. We also added causal attention masking to the SDPA layer for GPT-2. {\em DistilBERT}~\cite{sanh2019distilbert} is a knowledge-distilled version of BERT with fewer layers. {\em MobileBERT}~\cite{sun2020mobilebert} utilizes a bottleneck design and incorporates three additional feed-forward modules within each encoder, but with a smaller hidden dimension. {\em SqueezeBERT}~\cite{iandola2020squeezebert} retains the same structure as the {\em BERT-base} but replaces the linear kernels with convolutional kernels. 

\begin{figure}[t]
\scriptsize
\centering
\subfloat[CPU]{
\begin{tikzpicture}
\begin{axis}[
    width=\columnwidth, height=\chartheight*0.8, 
    ybar,
    bar width = .2cm,
    ymode = log,
    ymajorgrids=true, 
    yminorgrids=true,    
    enlargelimits=0.15,
    legend style={at={(0.5,-0.15)},
      anchor=north,legend columns=-1},
    ylabel={Latency [ms]},
    symbolic x coords={
                        {\em BERT-base}, 
                       {\em DistilBERT},
                       {\em MobileBERT},
                       {\em SqueezeBERT},
                       {\em GPT-2},
                       },
    xtick=data,
    xtick style={draw=none},
    xticklabel style={yshift=3pt},
    legend style={at={(0.5,1.05)}, font=\footnotesize, anchor=south, cells={anchor=west}, inner sep=0.2pt},
    legend columns=4,
    ]
\addplot [postaction={ pattern=north east lines},fill=green!20]
	coordinates {
            ({\em BERT-base},5209.438991000115)
            ({\em DistilBERT},2641.9417982)
            ({\em MobileBERT},334.03745852)
            ({\em SqueezeBERT},138.64243884)
            
            ({\em GPT-2},10957.647)
            };

\addplot [postaction={ pattern= north west lines},fill=yellow!20]
	coordinates {
            ({\em BERT-base},947.54)
            ({\em DistilBERT},728.14)
            ({\em MobileBERT},494.23)
            ({\em SqueezeBERT},337.77)
            ({\em GPT-2},1358.56)
            };

\addplot [postaction={ pattern=horizontal lines},fill=gray!20]
	coordinates {
            ({\em BERT-base},1662.18)
            ({\em DistilBERT},842.567)
            ({\em MobileBERT},324.228)
            ({\em SqueezeBERT},66.3513)

            ({\em GPT-2},2844.80)
            };

\legend{{\em TVM},{\em ExecuTorch}, {\em ARM-CL} (Extended)}
\end{axis}
\end{tikzpicture}
\label{fig:tvm_a}
}

\subfloat[GPU]{
\begin{tikzpicture}
\begin{axis}[
     width=\columnwidth, height=\chartheight *0.8, 
    ybar,
    bar width = .2cm,
    ymode = log,
    ymajorgrids=true, 
    yminorgrids=true,    
    enlargelimits=0.15,
    legend style={at={(0.5,-0.15)},
      anchor=north,legend columns=-1},
    ylabel={Latency [ms]},
    symbolic x coords={{\em BERT-base}, 
                       {\em DistilBERT},
                       {\em MobileBERT},
                       {\em SqueezeBERT},
                       {\em GPT-2},
                       },
    xtick=data,
    xticklabel style={yshift=3pt},
    xtick style={draw=none,},
    legend style={at={(0.5,1.05)}, font=\footnotesize, anchor=south, cells={anchor=west}, inner sep=0.2pt},
        legend columns=4,
    ]
\addplot [postaction={ pattern=north east lines},fill=green!20]
	coordinates {
            ({\em BERT-base},1033.117179)
            ({\em DistilBERT},514.5008556)
            ({\em MobileBERT},351.46461536)
            ({\em SqueezeBERT},490.6946314400004)

            ({\em GPT-2},2186.022)
            };

\addplot [postaction={ pattern= north west lines},fill=yellow!20]
	coordinates {
            ({\em BERT-base},0)
            ({\em DistilBERT},0)
            ({\em MobileBERT},0)
            ({\em SqueezeBERT},0)
            ({\em GPT-2},0)
            };

\addplot [postaction={ pattern=horizontal lines},fill=gray!20]
	coordinates {
            ({\em BERT-base},491.660)
            ({\em DistilBERT},244.237)
            ({\em MobileBERT},342.002)
            ({\em SqueezeBERT},130.738)

            ({\em GPT-2},579.1245)
            };

\end{axis}
\end{tikzpicture}
\label{fig:tvm_b}
}

\caption{Comparison between inference latency for different frameworks for different transformers. }
    \label{fig:tvm}
    \vspace{-0.5cm}
\end{figure}
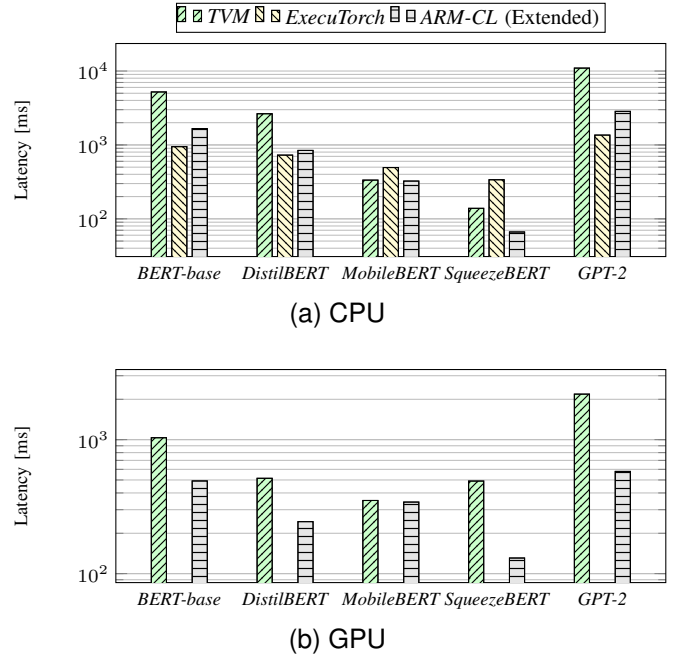

\textbf{Baseline Framework Comparison.} Figure~\ref{fig:tvm} compares the transformer inference latency for different transformers on unmodified {\em TVM v0.18} against our extended {\em ARM-CL}. Figures~\ref{fig:tvm_a} and~\ref{fig:tvm_b} show that our implementation reduces latency for CPU- and GPU-only transformer inference, compared to {\em TVM}, on average by 2.34x and 2.23x, respectively.
We also compare our extended {\em ARM-CL} with {\em ExecuTorch}, an edge-inference framework derived from {\em PyTorch}~\cite{imambi2021pytorch}. While {\em ExecuTorch} offers similar performance on the {\em ARM} CPU, it lacks support for the {\em ARM} GPU.

\textbf {Single-vs Multi-Processor Comparison.} Figure~\ref{fig:switching_results} illustrates the inference latency for both single- and multi-processor configurations of various transformers using our {\em ARM-CL} extensions. Within single-processor transformer inference using {\em ARM-CL}, the GPU outperforms the CPU for {\em BERT-base}, {\em DistilBERT}, and {\em GPT-2} while the CPU outperforms the GPU for {\em SqueezeBERT} and {\em MobileBERT}. However, the CPU-GPU layer-switched transformer inference proposed in this work outperforms CPU and GPU for all transformers. Overall, the CPU-GPU layer-switched transformer inference reduces the latency by up to 15.72\% and 10.95\% on average, compared to the best single-processor inference.

This work leverages the layer-wise computational heterogeneity of transformer models alongside hardware heterogeneity. Although GPU performance is bottlenecked by cache limitations for larger workloads, it can be improved with larger caches. Evaluated {\em Amlogic A331D} uses a 128 KB L2 cache, whereas the underlying {\em Bifrost} architecture supports up to 512\,KB, leaving room to benefit from layer-switched inference.  
Furthermore, the principle of assigning computational workloads to the most suitable processor to reduce overall inference latency can be generalized to other embedded HMPSoC platforms.

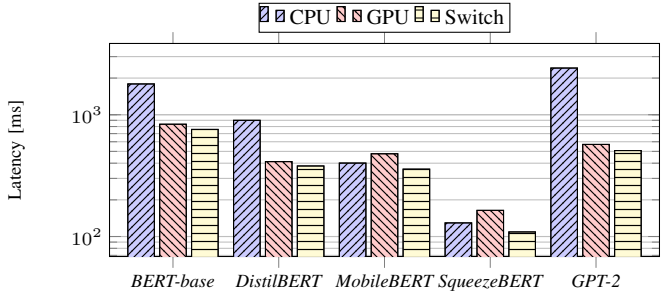
\begin{figure} [t]
    \centering
    \scriptsize
    \begin{tikzpicture}
\begin{axis}[
    width=\columnwidth, height=\chartheight * 0.8, 
    ybar,    
    ymode = log,
    ymajorgrids=true, 
    yminorgrids=true,
    enlargelimits=0.15,
    ylabel={Latency [ms]},
    symbolic x coords={{\em BERT-base}, 
                       {\em DistilBERT},
                       {\em MobileBERT},
                       {\em SqueezeBERT},
                       {\em GPT-2}},
    xtick=data,
    xticklabel style={rotate=0},
    legend style={at={(0.5,1.05)}, font=\footnotesize, anchor=south, cells={anchor=west}, inner sep=0.2pt},
        legend columns=4,
    ]
\addplot [postaction={ pattern=north east lines},fill=blue!20]
	coordinates {
            ({\em GPT-2},2417.39)
            ({\em BERT-base},1789.85)
            ({\em DistilBERT},899.57)
            ({\em MobileBERT},400.92)
            ({\em SqueezeBERT},129.38)
            };
\addplot [postaction={ pattern=north west lines},fill=red!20]
	coordinates {
            ({\em GPT-2},570.44)
            ({\em BERT-base},835.62)
            ({\em DistilBERT},412.11)
            ({\em MobileBERT},478.08)
            ({\em SqueezeBERT},163.97)
            };
\addplot [postaction={ pattern=horizontal lines},fill=yellow!20]
	coordinates {
            ({\em GPT-2},507.86)
            ({\em BERT-base},757.66)
            ({\em DistilBERT},379.62)
            ({\em MobileBERT},357.42)
            ({\em SqueezeBERT},109.06)
            };
\legend{CPU, GPU, Switch}
\end{axis}
\end{tikzpicture}
\vspace{-0.1cm}
\caption{Latency comparison between single- and multi-processor transformer inference for different transformers.}
    \label{fig:switching_results}
    \vspace{-0.3cm}
\end{figure}

\section {Conclusion}

We present extensions to {\em ARM-CL} that support the execution of different transformers on {\em ARM} CPUs and GPUs. The extensions provide a several-fold speedup in single-processor transformer inference on an {\em ARM}-based platform. We analyze the execution of transformer inference on various HMPSoC processors by utilizing micro-benchmarks. We present compelling evidence of performance variability in transformer layers on both embedded CPUs and GPUs, along with thorough reasoning for our findings. We present a case for cooperative CPU-GPU layer-switched on-chip low-latency transformer inference, making it feasible through a novel low-overhead implementation in {\em ARM-CL}. Our multi-processor approach decreases the inference latency of transformers compared to the single-processor method. 

\pagebreak

\bibliography{ref}

@inproceedings{taufique2025hidp,
  title={HiDP: Hierarchical DNN Partitioning for Distributed Inference on Heterogeneous Edge Platforms},
  author={Taufique, Zain and Vyas, Aman and Miele, Antonio and Liljeberg, Pasi and Kanduri, Anil},
  booktitle={2025 Design, Automation \& Test in Europe Conference (DATE)},
  pages={1--7},
  year={2025},
  organization={IEEE}
}

@inproceedings{taufique2025twill,
  title={Twill: Scheduling Compound AI Systems on Heterogeneous Mobile Edge Platforms},
  author={Taufique, Zain and Vyas, Aman and Miele, Antonio and Liljeberg, Pasi and Kanduri, Anil},
  booktitle={2025 IEEE/ACM International Conference On Computer Aided Design (ICCAD)},
  pages={1--9},
  year={2025},
  organization={IEEE}
}

@article{imambi2021pytorch,
  title={PyTorch},
  author={Imambi, Sagar and Prakash, Kolla Bhanu and Kanagachidambaresan, GR},
  journal={Programming with TensorFlow: solution for edge computing applications},
  pages={87--104},
  year={2021},
  publisher={Springer}
}

@article{sun2017enabling,
  title={Enabling embedded inference engine with arm compute library: A case study},
  author={Sun, Dawei and Liu, Shaoshan and Gaudiot, Jean-Luc},
  journal={arXiv preprint arXiv:1704.03751},
  year={2017}
}

@inproceedings{karatzas2023omniboost,
  title={Omniboost: Boosting throughput of heterogeneous embedded devices under multi-dnn workload},
  author={Karatzas, Andreas and Anagnostopoulos, Iraklis},
  booktitle={2023 60th ACM/IEEE Design Automation Conference (DAC)},
  year={2023},
  organization={IEEE}
}

@inproceedings{al2020novel,
  title={Novel casestudy and benchmarking of AlexNet for edge AI: From CPU and GPU to FPGA},
  author={Al-Ali, Firas and Gamage, Thilina Doremure and Nanayakkara, Hewa WTS and Mehdipour, Farhad and Ray, Sayan Kumar},
  booktitle={2020 IEEE Canadian Conference on Electrical and Computer Engineering (CCECE)},
  pages={1--4},
  year={2020},
  organization={IEEE}
}

@article{bahdanau2014neural,
  title={Neural machine translation by jointly learning to align and translate},
  author={Bahdanau, Dzmitry and Cho, Kyunghyun and Bengio, Yoshua},
  journal={arXiv preprint arXiv:1409.0473},
  year={2014}
}

@inproceedings{aghapour2022cpu,
  title={CPU-GPU layer-switched low latency CNN inference},
  author={Aghapour, Ehsan and Sapra, Dolly and Pimentel, Andy and Pathania, Anuj},
  booktitle={2022 25th Euromicro Conference on Digital System Design (DSD)},
  pages={324--331},
  year={2022},
  organization={IEEE}
}

@article{devlin2018bert,
  title={Bert: Pre-training of deep bidirectional transformers for language understanding},
  author={Devlin, Jacob and Chang, Ming-Wei and Lee, Kenton and Toutanova, Kristina},
  journal={arXiv preprint arXiv:1810.04805},
  year={2018}
}

@article{han2015deep,
  title={Deep compression: Compressing deep neural networks with pruning, trained quantization and huffman coding},
  author={Han, Song and Mao, Huizi and Dally, William J},
  journal={arXiv preprint arXiv:1510.00149},
  year={2015}
}

@inproceedings{chen2018tvm,
  title={$\{$TVM$\}$: An automated $\{$End-to-End$\}$ optimizing compiler for deep learning},
  author={Chen, Tianqi and Moreau, Thierry and Jiang, Ziheng and Zheng, Lianmin and Yan, Eddie and Shen, Haichen and Cowan, Meghan and Wang, Leyuan and Hu, Yuwei and Ceze, Luis and others},
  booktitle={13th USENIX Symposium on Operating Systems Design and Implementation (OSDI 18)},
  pages={578--594},
  year={2018}
}

@article{sun2020mobilebert,
  title={Mobilebert: a compact task-agnostic bert for resource-limited devices},
  author={Sun, Zhiqing and Yu, Hongkun and Song, Xiaodan and Liu, Renjie and Yang, Yiming and Zhou, Denny},
  journal={arXiv preprint arXiv:2004.02984},
  year={2020}
}

@article{iandola2020squeezebert,
  title={SqueezeBERT: What can computer vision teach NLP about efficient neural networks?},
  author={Iandola, Forrest N and Shaw, Albert E and Krishna, Ravi and Keutzer, Kurt W},
  journal={arXiv preprint arXiv:2006.11316},
  year={2020}
}

@article{sanh2019distilbert,
  title={DistilBERT, a distilled version of BERT: smaller, faster, cheaper and lighter},
  author={Sanh, Victor and Debut, Lysandre and Chaumond, Julien and Wolf, Thomas},
  journal={arXiv preprint arXiv:1910.01108},
  year={2019}
}

@article{chang2023pipebert,
  title={PipeBERT: high-throughput BERT inference for arm big. Little multi-core processors},
  author={Chang, Hung-Yang and Mozafari, Seyyed Hasan and Chen, Cheng and Clark, James J and Meyer, Brett H and Gross, Warren J},
  journal={Journal of Signal Processing Systems},
  volume={95},
  number={7},
  pages={877--894},
  year={2023},
  publisher={Springer}
}

@inproceedings{wang2022towards,
  title={Towards efficient vision transformer inference: A first study of transformers on mobile devices},
  author={Wang, Xudong and Zhang, Li Lyna and Wang, Yang and Yang, Mao},
  booktitle={Proceedings of the 23rd annual international workshop on mobile computing systems and applications},
  pages={1--7},
  year={2022}
}

@inproceedings{carion2020end,
  title={End-to-end object detection with transformers},
  author={Carion, Nicolas and Massa, Francisco and Synnaeve, Gabriel and Usunier, Nicolas and Kirillov, Alexander and Zagoruyko, Sergey},
  booktitle={European conference on computer vision},
  pages={213--229},
  year={2020},
  organization={Springer}
}

@article{garg2020bae,
  title={Bae: Bert-based adversarial examples for text classification},
  author={Garg, Siddhant and Ramakrishnan, Goutham},
  journal={arXiv preprint arXiv:2004.01970},
  year={2020}
}

@inproceedings{kim2020autoscale,
  title={Autoscale: Energy efficiency optimization for stochastic edge inference using reinforcement learning},
  author={Kim, Young Geun and Wu, Carole-Jean},
  booktitle={2020 53rd Annual IEEE/ACM international symposium on microarchitecture (MICRO)},
  pages={1082--1096},
  year={2020},
  organization={IEEE}
}

@inproceedings{dagli2024shared,
  title={Shared memory-contention-aware concurrent dnn execution for diversely heterogeneous system-on-chips},
  author={Dagli, Ismet and Belviranli, Mehmet E},
  booktitle={Proceedings of the 29th ACM SIGPLAN Annual Symposium on Principles and Practice of Parallel Programming},
  pages={243--256},
  year={2024}
}

@inproceedings{jeong2022band,
  title={Band: coordinated multi-dnn inference on heterogeneous mobile processors},
  author={Jeong, Joo Seong and Lee, Jingyu and Kim, Donghyun and Jeon, Changmin and Jeong, Changjin and Lee, Youngki and Chun, Byung-Gon},
  booktitle={Proceedings of the 20th Annual International Conference on Mobile Systems, Applications and Services},
  pages={235--247},
  year={2022}
}

@inproceedings{aghapour2024piqi,
  title={Piqi: Partially quantized dnn inference on hmpsocs},
  author={Aghapour, Ehsan and Shen, Yixian and Sapra, Dolly and Pimentel, Andy and Pathania, Anuj},
  booktitle={Proceedings of the 29th ACM/IEEE International Symposium on Low Power Electronics and Design},
  pages={1--6},
  year={2024}
}

@article{shen2026active,
  title={Active Imitation Learning for Thermal-and Kernel-Aware LFM Inference on 3D S-NUCA Many-Cores},
  author={Shen, Yixian and Shen, Chaoyao and Deen, Jan and Floros, George and Pimentel, Andy and Pathania, Anuj},
  journal={arXiv preprint arXiv:2604.11948},
  year={2026}
}

@article{li2026keeping,
  title={Keeping the Evidence Chain: Semantic Evidence Allocation for Training-Free Token Pruning in Video Temporal Grounding},
  author={Li, Jiaqi and Zheng, Shuntian and Shen, Yixian and Huang, Jia-Hong and Lu, Xiaoman and Ni, Minzhe and Guan, Yu},
  journal={arXiv preprint arXiv:2603.05663},
  year={2026}
}

@inproceedings{shen2025macp,
  title={MaCP: Minimal yet mighty adaptation via hierarchical cosine projection},
  author={Shen, Yixian and Bi, Qi and Huang, Jia-Hong and Zhu, Hongyi and Pimentel, Andy D and Pathania, Anuj},
  booktitle={Proceedings of the 63rd Annual Meeting of the Association for Computational Linguistics (Volume 1: Long Papers)},
  pages={20602--20618},
  year={2025}
}

@inproceedings{shen2025ssh,
  title={Ssh: Sparse spectrum adaptation via discrete hartley transformation},
  author={Shen, Yixian and Bi, Qi and Huang, Jia-Hong and Zhu, Hongyi and Pimentel, Andy D and Pathania, Anuj},
  booktitle={Proceedings of the 2025 Conference of the Nations of the Americas Chapter of the Association for Computational Linguistics: Human Language Technologies (Volume 1: Long Papers)},
  pages={10400--10415},
  year={2025}
}

@inproceedings{shen2022tcps,
  title={TCPS: a task and cache-aware partitioned scheduler for hard real-time multi-core systems},
  author={Shen, Yixian and Xiao, Jun and Pimentel, Andy D},
  booktitle={Proceedings of the 23rd ACM SIGPLAN/SIGBED International Conference on Languages, Compilers, and Tools for Embedded Systems},
  pages={37--49},
  year={2022}
}

@article{xiao2022cache,
  title={Cache interference-aware task partitioning for non-preemptive real-time multi-core systems},
  author={Xiao, Jun and Shen, Yixian and Pimentel, Andy D},
  journal={ACM Transactions on Embedded Computing Systems (TECS)},
  volume={21},
  number={3},
  pages={1--28},
  year={2022},
  publisher={ACM New York, NY}
}

@article{shen2023thermal,
  title={Thermal management for 3d-stacked systems via unified core-memory power regulation},
  author={Shen, Yixian and Schreuders, Leo and Pathania, Anuj and Pimentel, Andy D},
  journal={ACM Transactions on Embedded Computing Systems},
  volume={22},
  number={5s},
  pages={1--26},
  year={2023},
  publisher={ACM New York, NY}
}
\bibliographystyle{IEEEtran}

\end{document}